\documentclass[10pt,twocolumn,superscriptaddress,showpacs,preprintnumbers,amsmath,amssymb,prl,aps]{revtex4}
\usepackage{graphicx}

\usepackage{ulem}   
\begin{document}


\title{Evidence for Time-Reversal-Symmetry-Broken Superconductivity in Locally Noncentrosymmetric SrPtAs}

\author{P. K. Biswas}
\affiliation{Laboratory for Muon Spin Spectroscopy, Paul Scherrer Institute, CH-5232 Villigen PSI, Switzerland}

\author{H. Luetkens}
\email[]{hubertus.luetkens@psi.ch}
\affiliation{Laboratory for Muon Spin Spectroscopy, Paul Scherrer Institute, CH-5232 Villigen PSI, Switzerland}

\author{T. Neupert}
\affiliation{Condensed Matter Theory Group, Paul Scherrer Institute, CH-5232 Villigen PSI, Switzerland}
\affiliation{Institute for Theoretical Physics, ETH Zurich, 8093 Zurich, Switzerland}

\author{T. St\"{u}rzer}
\affiliation{Department Chemie, Ludwig-Maximilians-Universit\"{a}t M\"{u}nchen, D-81377 M\"{u}nchen, Germany}

\author{C. Baines}
\affiliation{Laboratory for Muon Spin Spectroscopy, Paul Scherrer
Institute, CH-5232 Villigen PSI, Switzerland}

\author{G. Pascua}
\affiliation{Laboratory for Muon Spin Spectroscopy, Paul Scherrer
Institute, CH-5232 Villigen PSI, Switzerland}

\author{A. P. Schnyder}
\affiliation{Max-Planck-Institut f\"{u}r Festk\"{o}rperforschung, Heisenbergstrasse 1, D-70569 Stuttgart, Germany}

\author{M. H. Fischer}
\affiliation{Department of Physics, Cornell University, Ithaca, New York 14853, USA}

\author{J. Goryo}
\affiliation{Institute of Industrial Science, The University of Tokyo, Meguro, Tokyo 153-0041, Japan}
\affiliation{Institute for Theoretical Physics, ETH Zurich, 8093 Zurich, Switzerland}

\author{M. R. Lees}
\affiliation{Physics Department, University of Warwick, Coventry,
CV4 7AL, United Kingdom}

\author{H. Maeter}
\affiliation{Institute for Solid State Physics, TU Dresden,
D-01069 Dresden, Germany}

\author{F. Br\"uckner}
\affiliation{Institute for Solid State Physics, TU Dresden,
D-01069 Dresden, Germany}
\author{H.-H. Klauss}
\affiliation{Institute for Solid State Physics, TU Dresden, D-01069 Dresden, Germany}

\author{M. Nicklas}
\affiliation{Max Planck Institute for Chemical Physics of Solids,
N\"othnitzer Str. 40, 01187 Dresden, Germany}

\author{P. J. Baker}
\affiliation{ISIS Facility, STFC Rutherford Appleton Laboratory,
Didcot OX11 0QX, United Kingdom}

\author{A. D. Hillier}
\affiliation{ISIS Facility, STFC Rutherford Appleton Laboratory,
Didcot OX11 0QX, United Kingdom}

\author{M. Sigrist}
\affiliation{Institute for Theoretical Physics, ETH Zurich, 8093 Zurich, Switzerland}

\author{A. Amato}
\affiliation{Laboratory for Muon Spin Spectroscopy, Paul Scherrer Institute, CH-5232 Villigen PSI, Switzerland}

\author{D. Johrendt}
\affiliation{Department Chemie, Ludwig-Maximilians-Universit\"{a}t M\"{u}nchen, D-81377 M\"{u}nchen, Germany}

\date{\today}

\begin{abstract}
We report the magnetic and superconducting properties of locally
noncentrosymmetric SrPtAs obtained by
muon-spin-rotation/relaxation ($\mu$SR) measurements. Zero-field
$\mu$SR reveals the occurrence of small spontaneous static
magnetic fields with the onset of superconductivity. This finding
suggests that the superconducting state of SrPtAs breaks
time-reversal symmetry. The superfluid density as determined by
transverse field $\mu$SR is nearly flat approaching $T=0$~K
proving the absence of extended nodes in the gap function. By
symmetry, several superconducting states supporting time-reversal
symmetry breaking in SrPtAs are allowed. Out of these, a
dominantly $d + id$ (chiral $d$-wave) order parameter is most
consistent with our experimental data.
\end{abstract}



\maketitle

Transition metal pnictides have attracted considerable scientific
interest as they present the second largest family of
superconductors after the cuprates~\cite{Kamihara}. All
superconductors of this family share one common structural
feature: superconductivity takes place in a square lattice formed
by the transition metal elements. Very recently superconductivity
with a $T_c$ of 2.4~K has been discovered in
SrPtAs~\cite{Nishikubo}, which has a unique and attractive
structural feature: It crystallizes in a hexagonal structure with
weakly coupled PtAs layers forming a honeycomb lattice. SrPtAs
supports three pairs of split Fermi surfaces, two of which are
hole-like and centered around the $\Gamma$-point with a
cylindrical shape extended along the $k_z$ direction and together
host only about 30$\%$ of the density of states. The remaining
70$\%$ of the density of states are hosted by the third pair of
split Fermi surfaces that is electron-like, centered around the
$K$ and $K'$ and consists of a cylindrical and a cigar-like
sheet~\cite{Youn, Shein}. One unit cell of SrPtAs contains two
PtAs layers each of which lacks a center of inversion symmetry
even though the system has a global inversion center~\cite{Youn}.
Locally broken inversion symmetry in SrPtAs together with a strong
spin-orbit coupling might cause dramatic effects on the
superconducting properties of this system that are otherwise found
in noncentrosymmetric materials only \cite{Fischer11}. Indeed,
theoretical calculations focusing on a spin-singlet order
parameter for SrPtAs predict a significant enhancement of the
Pauli limiting field and the zero-temperature spin susceptibility
\cite{Youn}. In addition, a comprehensive symmetry analysis
reveals that some unconventional states are possible, such as the
$A_{2u}$ state with a dominant $f$-wave component and the $E_g$
state with a dominant chiral \textit{d}-wave part, which breaks
time-reversal symmetry (TRS)~\cite{Goryo}.


In this Letter, we report on muon spin-rotation/relaxation
($\mu$SR) measurements to determine the magnetic and
superconducting properties of SrPtAs. We find small spontaneous
internal magnetic fields below $T_c$ showing that the
superconducting state breaks TRS. Low-temperature superfluid
density measurements indicate the absence of extended nodes in the
gap function of SrPtAs. These experimental findings are discussed
in light of the different superconducting states allowed by
symmetry. From these states, the $E_g$ (chiral $d$-wave) order
parameter is the most likely pairing state in SrPtAs. We also
discuss some other possible scenarios.

Two batches of polycrystalline samples (A and B) of SrPtAs were
prepared via a solid state reaction method as described in
Ref.~\cite{Nishikubo}. Sample A is a disk-shaped pellet
($\approx12$~mm diameter and 1~mm thickness), while sample B is a
powder of polycrystalline SrPtAs. Both samples were glued to a Ag
sample holder. Low-temperature $\mu$SR measurements on sample A
were carried out down to 0.019~K using the low-temperature
facility (LTF) muon instrument located on the $\pi$M3 beamline of
the Swiss Muon Source at the Paul Scherrer Institute, Villigen,
Switzerland. Analogous measurements were carried out on sample B
using the MuSR spectrometer at the ISIS pulsed muon facility,
Oxford, United Kingdom. Data were collected with zero (ZF),
longitudinal (LF), and transverse magnetic fields (TF). The
magnetic field was applied above the superconducting transition
temperature and the sample subsequently cooled down to base
temperature.

\begin{figure}[tb!]
\begin{center}
\includegraphics[width=1.0\columnwidth]{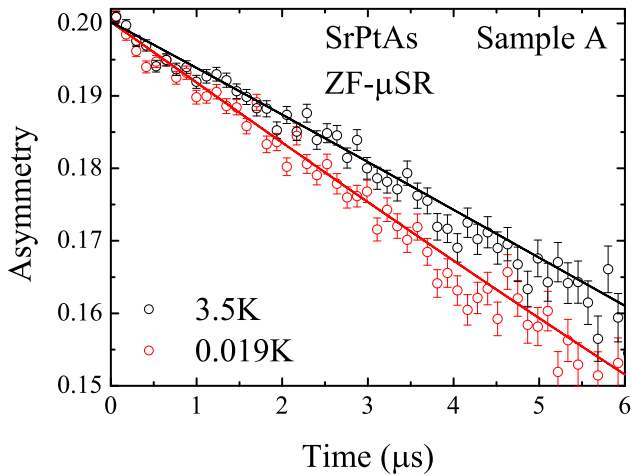}
\caption{\label{Figure1Biswas} (Color online) ZF-$\mu$SR time
spectra for SrPtAs at 3.5~K (above $T_c$, open circles) and
0.019~K (below $T_c$. closed circles). The solid lines are the
fits to the data using Eq.~(\ref{ZF_Fit}).}
\end{center}
\end{figure}

\begin{figure}[tb!]
\begin{center}\includegraphics[width=1.0\columnwidth]{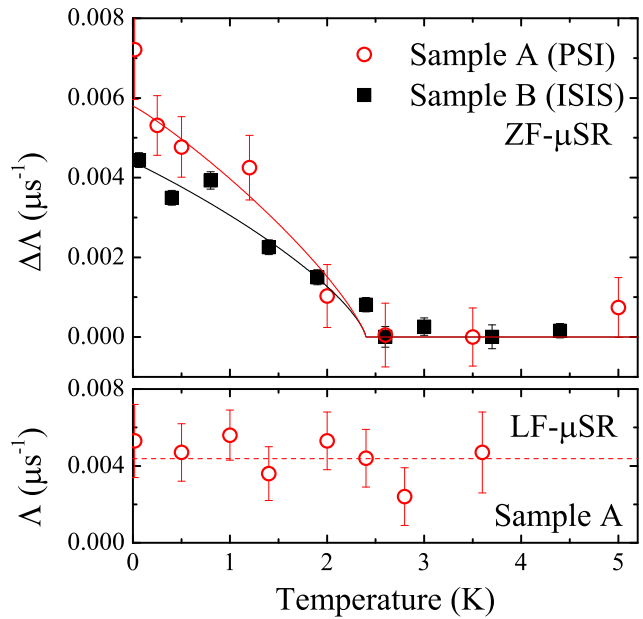}
\caption{\label{Figure2Biswas} (Color online) Upper panel:
Temperature-dependent part of the electronic relaxation rate
$\Delta\Lambda$ for two different samples of SrPtAs measured at
two different $\mu$SR facilities. Lower panel: Temperature
dependence of $\Lambda$ (open circles) in a weak LF of 9~mT. Solid
curves are guides to the eye.}
\end{center}
\end{figure}

Figure~\ref{Figure1Biswas} shows ZF-$\mu$SR time spectra for
SrPtAs. Open circles indicate data collected at 3.5~K (above
$T_c$) and closed circles data collected at 0.019~K (below $T_c$).
Data taken below $T_c$ show a greater relaxation than above $T_c$.
The small relaxation above $T_c$ arises from randomly orientated
nuclear magnetic dipole moments that are static on the time scale
of $\mu$SR. The additional relaxation of the muon spin
polarization $P(t)$ below $T_c$ is caused by spontaneous magnetic
moments which may be either quasi-static or dynamic. To
distinguish between these two possibilities, we have performed
LF-$\mu$SR in a weak magnetic field of 9~mT. We find that this
field is sufficient to decouple the muon spin polarization (see
the lower panel of Fig.~\ref{Figure2Biswas}) which proves that the
additional spontaneous magnetic relaxation appearing below $T_c$
in the ZF-$\mu$SR data is due to quasi-static moments
\cite{Yaouanc11}. To quantitatively evaluate the ZF-$\mu$SR
spectra, we fitted a combination of a static Lorentzian and
Gaussian Kubo-Toyabe relaxation
function~\cite{Kubo,Hayano,Uemura,Maisuradze10} to the data:

\begin{equation}\label{ZF_Fit}
P(t)=\frac{1}{3}+\frac{2}{3}(1-\Delta_{nm}^{2}t^{2}-\Lambda{t})\exp\left(-\frac{\Delta_{nm}^{2}t^{2}}{2}-\Lambda{t}\right)\;.
\end{equation}
Here $\Delta_{nm}$ and $\Lambda$ are the temperature-independent
Gaussian and the temperature-dependent exponential muon relaxation
rates due to the presence of nuclear and electronic moments,
respectively. The solid lines in Fig.~\ref{Figure1Biswas} are
fits to the data using this equation.

Both samples of SrPtAs measured at two different $\mu$SR
facilities show very similar spectra. In both cases a small
increase of the relaxation rate is observed below the
superconducting $T_c$. Figure~\ref{Figure2Biswas} shows the
temperature-dependent part of the relaxation rate
$\Delta\Lambda=\Lambda(T) - \Lambda(T\approx 5\,\mathrm{K})$ for
both samples. The increase of the relaxation indicates the
appearence of a spontaneous magnetic field in the superconducting
state of SrPtAs. The existence of such a spontaneous field in
SrPtAs and its correlation with the superconducting $T_c$ provides
evidence for a superconducting state that breaks TRS. A similar
behavior has been observed by $\mu$SR in the spin-triplet
TRS-breaking superconductor Sr$_2$RuO$_4$~\cite{Luke98}. Possible
origins for the occurrence of TRS breaking in SrPtAs will be
discussed below, but we first turn to the experimental
characterization of the superconducting properties of SrPtAs.

\begin{figure}[tb!]
\begin{center}\includegraphics[width=1.0\columnwidth]{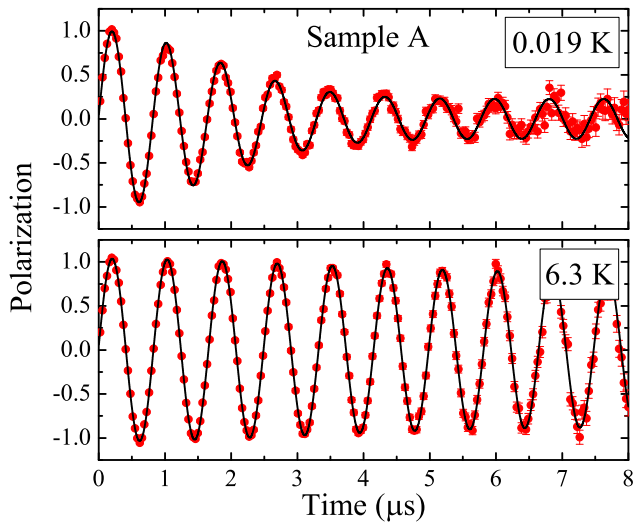}
\caption{\label{Figure3Biswas} (Color online) TF-$\mu$SR time
spectra of SrPtAs collected at 0.019 and 6.3~K. Solid lines are
fits to the data using Eq.~(\ref{Depolarization_Fit}).}
\end{center}
\end{figure}

To reveal information on the pairing symmetry in the presence of
TRS breaking, we have also performed TF-$\mu$SR measurements in a
field of 9~mT, which is larger than the first critical field of
$\approx 4$~mT that we determined by low temperature magnetization
measurements. Figure~\ref{Figure3Biswas} shows TF-$\mu$SR
precession signals of sample A of SrPtAs above and below $T_c$. In
the normal state, the oscillation only shows very small
relaxation. Below $T_c$, the relaxation rate increases due to the
broad field distribution produced by the presence of the vortex
lattice. Solid lines are fits to the data using a sinusoidally
oscillating function with a Gaussian decay component:

\begin{eqnarray}
\label{Depolarization_Fit}
\begin{split}
P(t)= &\; P_{s}\exp\left(-\sigma^{2}t^{2}\right/2)\cos\left(\gamma_\mu B_{int}t +\phi\right)\\
&+P_{bgd}\cos\left(\gamma_\mu B_{bgd}t +\phi\right),~
\end{split}
\end{eqnarray}
where $P_s$ and $P_{bgd}$ are the relative fractions of muons
hitting the sample and the Ag sample holder, respectively. The
latter giving a practically undamped background signal.
$\gamma_{\mu}/2\pi=135.5$~MHz/T is the muon gyromagnetic
ratio~\cite{Sonier}, $B_{int}$ and $B_{bgd}$ are the internal and
background magnetic field at the muon sites, $\phi = -\pi/2$ is
the phase of the initial muon spin polarization with respect to
the positron detector and $\sigma$ is the Gaussian muon spin
relaxation rate. $\sigma$ can be written as
$\sigma=\left(\sigma^{2}_{sc} +
\sigma^{2}_{nm}\right)^{\frac{1}{2}}$, where $\sigma_{sc}$ is the
superconducting contribution to the relaxation rate due to the
field variation across the flux line lattice, and $\sigma_{nm}$ is
the nuclear magnetic dipolar contribution which is assumed to be
constant over the temperature range of the study.

\begin{figure}[tb!]
\begin{center}
\includegraphics[width=1.0\columnwidth]{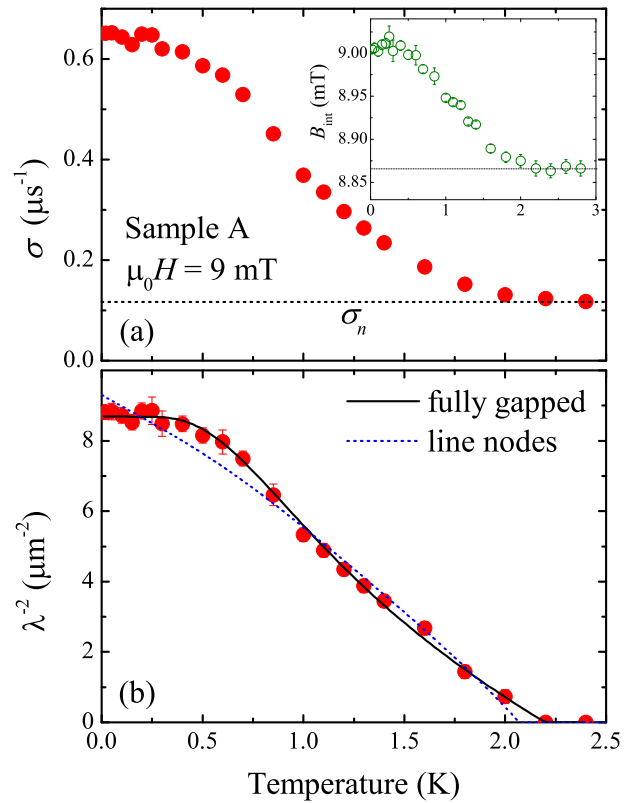}
\caption{\label{Figure4Biswas} (Color online) (a) Temperature
dependence of $\sigma_{sc}$ of SrPtAs. The inset is the
temperature dependence of the internal field. (b) Temperature
dependence of $\lambda^{-2}$ of SrPtAs which is proportional to
the superfluid density $n_s/m^\ast$. Solid and dashed lines are
the fits to the data using a fully gapped model and a one-band
model with line nodes, respectively.}
\end{center}
\end{figure}

Figure~\ref{Figure4Biswas}(a) shows the temperature dependence of
$\sigma$ for SrPtAs for sample A. Qualitatively similar data are
obtained for sample B. In a superconductor with an ideal
Ginzburg-Landau vortex lattice, $\sigma_{sc}$ is related to the
penetration depth $\lambda$ via Brandt's equation~\cite{Brandt}
\begin{equation}\label{sigma_lambda}
\sigma_{sc}=4.83\times10^{4}(1-b)[1+1.21(1-\sqrt{b})^3]\lambda^{-2},
\end{equation}
where $b=H/H_{c2}$ is the ratio of the applied field to the upper
critical field. The temperature dependence of $\lambda^{-2}$ is
shown in Fig.~\ref{Figure4Biswas}(b) for which we used the data of
Ref.~\cite{Nishikubo} to estimate the temperature dependence of
$H_{c2}$. The slope of the temperature dependence of the
superconducting order parameter tends to zero as the temperature
approaches zero, indicating the absence of low-lying excitations.
This is corroborated by a reasonable fit of the data with a
one-band BCS \textit{s}-wave model. In contrast, a one-band model
with line nodes is unable to reproduce the data~\footnote{For the
displayed fit with line nodes, we use a one band \textit{d}-wave
model. Details of models are discussed in
Refs.~\cite{Carrington,Padamsee} and also in our previous
work~\cite{Biswas,Khasanov}.}. However, we emphasize that in
SrPtAs multiple electronic bands cross the Fermi level and that
SrPtAs lacks locally an inversion symmetry center. Hence,
unconventional order parameter symmetries and multiband
superconductivity can be expected. Therefore, such a simple model
description of the superfluid density is not likely to reveal the
true underlying pairing symmetry. However, with the present
statistical accuracy we are at least confident that SrPtAs has no
extended nodes in the gap function. This conclusion holds for both
samples even though the measured low-temperature magnetic
penetration depth is about 30\% smaller for sample B
($\lambda=239(4)$~nm) than for sample A ($\lambda=339(6)$~nm). The
reason for this discrepancy is not known at the moment.

Interestingly, the internal magnetic field $B_{int}$ clearly
increases by about 0.15~mT below $T_c$ as shown for sample A in
the inset of Fig.~\ref{Figure4Biswas}. This behavior is unexpected
since typically one observes a diamagnetic shift in spin-singlet
superconductors or a constant field in spin-triplet
superconductors. More detailed studies, especially on single
crystals, for which demagnetization effects can be properly taken
into account are required to understand this phenomenon in detail,
but we suggest that this positive internal field shift might be
related to the occurrence of TRS breaking in SrPtAs.

In summary, we found experimental evidence for (i) weak breaking
of TRS that sets in at the same temperature as the superconducting
order and (ii) no extended nodes in the order parameter. In the
following, we will discuss possible origins of TRS breaking for
different order parameter symmetries in the light of these
experimental observations \footnote{A microscopic analysis of
time-reversal-symmetry breaking will be presented elsewhere: M. H.
Fischer \textit{et al.}, in preparation.}.

\begin{figure}[t]
\begin{center}
\includegraphics[width=0.9\columnwidth]{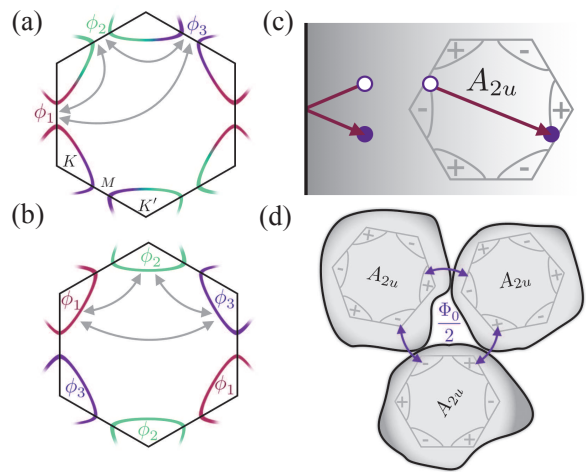}
\caption{ (Color online) Illustration of possible mechanisms for
TRS breaking in SrPtAs.
(a) Frustrated inter-Fermi surface Cooper pair scattering.
(b) Stabilizing the chiral $E^{\ }_g$ state with Cooper pair scattering.
(c) Andreev scattering with $\pi$ phase
shift in the $A^{\ }_{2u}$ state.
(d) Frustrated current loops.
}
\label{fig: Mechanisms}
\end{center}
\end{figure}

\emph{i) Chiral bulk and surface states}
---
The most straight-forward explanation for
bulk TRS breaking is a superconducting order parameter of dominantly $d+id$ symmetry, the so-called chiral $E^{\ }_{g}$ state~\cite{Goryo}.
This state can be stabilized in a scenario where superconductivity
is driven by the van Hove singularities located around the
$M$-points, see Fig.~\ref{fig: Mechanisms}. Intuitively, this can
be explained using the following parallel to the cuprates: On the
square lattice, repulsive scattering between the \emph{two} van
Hove singularities favors the $d_{x^2-y^2}$ order parameter
symmetry with \emph{sign change}. On a lattice with three-fold
rotational symmetry, repulsive scattering between the \emph{three}
van Hove singularities is frustrated. The phases of the
superconducting order parameter near the van Hove points,
$\phi_i,\ i=1,2,3$, then spontaneously choose one of the TRS
breaking \emph{chiral} configurations $\phi^{\ }_{1}-\phi^{\
}_{2}=\phi^{\ }_{2}-\phi^{\ }_{3}=\pm2\pi/3$ [see Fig.~\ref{fig:
Mechanisms}(a)]~\cite{Nandkishore12,Kiesel12}. The $E^{\ }_{g}$
state supports topologically protected chiral modes at the surface
and at defects, such as the implanted muon itself.

\emph{ii) Frustrated interband Cooper pair scattering}
---
An alternative scenario for bulk TRS breaking is based on frustrated Cooper pair scattering between multiple disconnected
Fermi pockets~\cite{Voelker02} located at the $K$ and $K'$ points.
Starting from a nodeless $s$-wave superconducting order parameter ($A_{1g}$ symmetry),
this inter-pocket scattering leads to a similar chiral phase structure as in scenario i),
however, with constant phases $\phi_i,\ i=1,2,3$, on each pair of pockets [see Fig.~\ref{fig: Mechanisms}(b)].
Here, the TRS breaking is a secondary transition out of a conventional $s$-wave superconductor.

\emph{iii) TRS breaking at the surface}
---
In view of the granularity of our samples, we comment on the possibility of spontaneous TRS breaking at the boundary, while the bulk remains in a TRS superconducting state.
Such surface effects can be expected
for a dominantly $f$-wave superconductor ($A^{\ }_{2u}$ state),
with a gap function that changes sign under a $\pi/3$ rotation around the $z$ axis [see Fig.~\ref{fig: Mechanisms}(c)].
On the one hand, this sign change influences the Andreev scattering on the sample surface, if the surface normal has a nonvanishing component in the plane
defined by the layers [see Fig.~\ref{fig: Mechanisms}(c)].
Energetically, the destructively interfering scattering process
leads to a suppression of the $A^{\ }_{2u}$ state close to the
surface. Topologically, it equips the surface with a
dispersionless 2D band structure with a degeneracy protected by
TRS. Residual interactions in this surface flat  band  generically
lift its degeneracy, thereby breaking spontaneously the TRS
by nucleation of a
$\Delta_{A_{2u}}+\mathrm{i}\Delta_{A_{1g}}$ order parameter locally at the surface.
On the other hand, due to the granularity of the sample, three or
more crystallites can form a loop around a void within the sample,
while remaining in a phase coherent superconducting state. Due to
the sign change of the gap function in the $A^{\ }_{2u}$ state,
the order parameter might acquire a phase shift of $\pi$ around
the loop~\cite{Sigrist-Rice}. To account for this phase shift,
half of a superconducting flux quantum $\phi_0/2=h/(4e)$ is
trapped in the loop [see Fig.~\ref{fig: Mechanisms}(d)]. The
associated TRS breaking supercurrent would then cause an increase
in muon depolarization rate of the muons that stop close to the
loop. At low fields these frustrated current loops are expected to
give rise to an unusual positive field shift, an effect that has
also been observed in granular samples of  high-$T_c$
cuprates~\cite{Sigrist-Rice, Braunish}. This granularity scenario
would be consistent with the positive field shift below $T_c$ of
the sample (see inset of Fig.~\ref{Figure4Biswas}).

Notably, none of the above scenarios is compatible with an $s$-wave
order parameter symmetry. Furthermore, in the absence of
fine tuning, TRS breaking in scenarios ii) and iii)  is expected to
occur as a secondary transition, i.e., at a temperature lower than
$T_c = 2.4$ K, which is in contradiction with our measurements.
Hence, the chiral $E^{\ }_{g}$ state is the most likely
explanation for TRS breaking in SrPtAs. This state is also favored
by the second experimental finding that the gap function of SrPtAs
possesses no extended nodes since it has only a point node on one
out of six bands contributing to superconductivity, while the
other pairing symmetry allowing for TRS breaking, namely $A^{\
}_{2u}$, possesses line nodes on four of these bands.

In conclusion, we have found small static magnetic fields in two
different samples of SrPtAs measured at two different $\mu$SR
facilities. These internal magnetic fields set in at the
superconducting $T_c$ and grow with decreasing temperature. This
provides strong evidence that the superconducting state of locally
noncentrosymmetric SrPtAs breaks TRS. In addition, we observed an
nearly flat temperature dependence of the superfluid density
approaching $T=0$~K proving the absence of extended nodes in the
superconducting gap function. Furthermore we proposed several
scenarios which can lead to TRS breaking in SrPtAs on the basis of
the superconducting states allowed by symmetry. While the other
states can not be completely excluded, our experimental
observations are most consistent with an $E^{\ }_{g}$ symmetry of
the superconducting order parameter which is dominated by a
spin-singlet $d + id$ (chiral $d$-wave) state.

The $\mu$SR experiments were performed at the Swiss Muon Source,
Paul Scherrer Institut, Villigen, Switzerland and the ISIS pulsed
muon facility, Oxford, United Kingdom. MHF acknowledges support
from NSF Grant DMR-0955822 and from NSF Grant DMR-1120296 to the
Cornell Center for Materials Research.

\bibliography{Biswas}

\end{document}